\documentclass[apj]{emulateapj}
\usepackage{graphicx} 
\usepackage[T1]{fontenc} 

\begin{document}
\title{Properties of a Gamma Ray Burst Host Galaxy at $z\sim5$
\altaffilmark{1}}
\author{
P.~A.~Price,$\!$\altaffilmark{2} 
A.~Songaila,$\!$\altaffilmark{2} 
L.~L.~Cowie,$\!$\altaffilmark{2} 
J.~Bell Burnell,$\!$\altaffilmark{3}
E.~Berger,$\!$\altaffilmark{4} 
A.~Cucchiara,$\!$\altaffilmark{5} 
D.~B.~Fox,$\!$\altaffilmark{5} 
I.~Hook,$\!$\altaffilmark{3} 
S.~R.~Kulkarni,$\!$\altaffilmark{6}
B.~Penprase,$\!$\altaffilmark{7} 
K.~C.~Roth,$\!$\altaffilmark{8} 
B.~Schmidt,$\!$\altaffilmark{9} 
}

\altaffiltext{1}{Based on observations obtained at the Gemini Observatory,
which is operated by the Association of Universities for Research in
Astronomy, Inc., under a cooperative agreement with the NSF on behalf of the
Gemini Partnership:  The National Science Foundation (United States), the
Particle Physics and Astronomy Research Council (United Kingdom), the National
Research Council (Canada), CONICYT (Chile), the Australian Research Council
(Australia), CNPq (Brazil) and CONICET (Argentina)}
\altaffiltext{2}{Institute for Astronomy, University of Hawaii,
   2680 Woodlawn Drive, Honolulu, HI 96822}
\altaffiltext{3}{Astrophysics, University of Oxford, Denys Wilkinson Building,
Keble Road, Oxford OX1 3RH, UK}
\altaffiltext{4}{Observatories of the Carnegie Institution of Washington, 813
Santa Barbara Street Pasadena CA 91101}
\altaffiltext{5}{Dept. of Astronomy and Astrophysics, Pennsylvania State
University, 525 Davey Laboratory University Park, PA 16802}
\altaffiltext{6}{Mail Code 105-24 Astronomy, Caltech, 1200 East California
Blvd., Pasadena, CA 91125}
\altaffiltext{7}{Pomona College 610 North College Avenue,
Claremont CA 91711}
\altaffiltext{8}{Gemini Observatory, 670 N. A'ohoku St., Hilo, HI 96720}
\altaffiltext{9}{Research School of Astronomy and Astrophysics, Australian National University, Mount Stromlo Observatory, Cotter Road, Weston Creek, Canberra ACT 2611, Australia}

\shorttitle{$z\sim5$ GRB host galaxy}
\shortauthors{Price et al.\/}


\begin{abstract}

We describe the properties of the host galaxy of the gamma-ray
burst GRB060510B based on a spectrum of the burst afterglow
obtained with the Gemini North 8m telescope. The galaxy lies at
a redshift of $z=4.941$ making it the fourth highest spectroscopically
identified burst host. However, it is the second highest redshift galaxy
for which the quality of the spectrum permits a detailed metallicity
analysis. The neutral hydrogen column density has a logarithmic
value of 21.0--21.2 ${\rm cm}^{-2}$ and the weak metal lines of Ni, S and
Fe show that the metallicity is in excess of a tenth of solar
which is far above the metallicities in damped Lyman alpha 
absorbers at high redshift. The tightest constraint is from
the Fe lines which place [Fe/H] in excess of  $-0.8$.
We argue that the results suggest that metallicity bias
could be a serious problem with inferring star formation from
the GRB population and consider how future higher quality measurements
could be used to resolve this question.

\end{abstract}


\section{Introduction}
\label{secintro}

Long duration gamma-ray bursts are believed to form in the collapse of massive
stars (e.g. Woosley 1993, Stanek et al. 2003) and hence to be a tracer of
galaxies with ongoing star formation (Lamb and Reichert 2000).  The afterglows
associated with the burst are also extremely bright immediately post-burst and
can be seen to very large redshifts, currently out to $z=6.2$ (Kawai et
al. 2006). Prompt spectroscopy of the afterglow can therefore allow us to
obtain the redshift of the galaxy in which the GRB has occurred and to study
the metallicity of the interstellar medium in this host galaxy.

With the caveat that we still do not fully understand the selection biases
introduced by the formation of the GRBs, this may be the best 
way in which we can study the properties of individual galaxies at very high
redshifts ($z>5$) in detail.  
The galaxies themselves are too faint (magnitudes of about
25) for detailed spectroscopic study, nor can they be identified as damped
L$\alpha$ absorbers (DLAs) at these redshifts since the L$\alpha$ forest
becomes too thick to allow us to identify the DLAs and measure their column
densities.

However, in order to take advantage of GRBs to study the high redshift galaxy
population we must be able both to localize the GRB quickly and then to
rapidly identify it as a candidate high redshift object. The $\it{Swift}$
satellite launched in 2004 (Gehrels et al. 2004) has made such studies
possible by providing a large sample of GRBs with accurate positions to faint
gamma ray detection thresholds. At the $\it{Swift}$ limits Jakobson et
al. (2006a) estimate that about 7\% of the GRBs are at $z>5$ so that there
should be a handful of such objects available for study in each year of the
$\it{Swift}$ mission.

Given this small number of available targets it is critical to observe as many
as possible with the highest quality and highest resolution spectroscopy that
we can obtain.  However, even with the $\it{Swift}$ data this remains
extremely challenging since we must first identify the afterglow and since the
subsequent spectroscopy requires target of opportunity obervations (TOOs) on
the largest ground based telescopes. Fortunately, some of the 8m class
telescopes are operated in queue mode and can respond rapidly to such TOO
events.  The present program utilizes this capability on the two Gemini 8m
telescopes where we can initiate observations with the GMOS spectrographs
(Hook et al. 2004) almost immediately when a candidate high redshift burst is
identified. The GMOS spectral resolution (R$\sim2000$\ in the mode which we
used) is not as high as desirable but is adequate for measuring the redshift
and providing metallicity estimates if we can obtain high signal-to-noise
spectra.

This {\it Letter\/} describes our GMOS observations of GRB060510B
which we find to lie at a redshift of $z=4.941$. This is the fourth
highest redshift identified for a GRB (e.g. Jakobson et al. (2006b)
for a recent summary) but is the second highest for which we can make
a metallicity analysis. The highest redshift analysis is of the
$z=6.2$ GRB (Kawai et al. 2006). We find that the metallicity is in
excess of a tenth solar, which is much higher than is seen in DLAs at
$z>4$ and also suggests that the metallicity of the GRB host galaxies
is not changing as a function of redshift.

\section{Observations}
\label{secobs}

  GRB060510B was detected by $\it{Swift}$ on May 10th 2006 at 8:22:24
UT. The XRT on $\it{Swift}$ began observing the area 119s after the trigger and a
bright X-ray source was located at RA=15 56 29.3 and Dec=78 34 09.4
(J2000). The long duration gamma-ray burst was noted as a potential high redshift
object (Krimm et al. 2006). A fading optical counterpart was detected
by Mirabal and Halpern (2006)  at RA=15 56 29.615 and Dec=78 34 13.02. The red
(R-I) color was again suggestive of a possible high redshift candidate. NIRI
observations with Gemini-North detected the counterpart with J$\sim$19 
at just over 2 hours after the trigger (Price et al. 2006a). 

\begin{figure}
\begin{center}
 \includegraphics[viewport=20 0 446
 626,width=2.4in,clip,angle=90,scale=0.95]{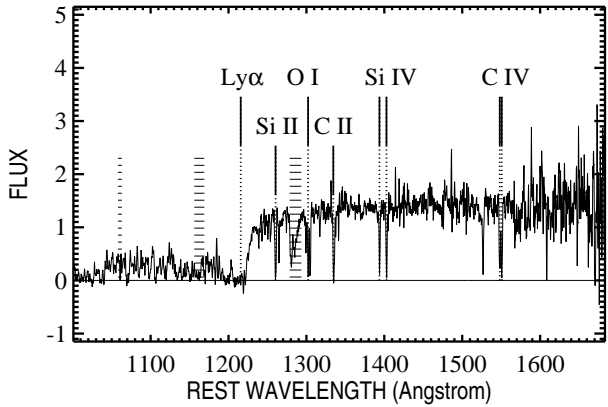}
 \caption{Spectrum of GRB060510B obtained with the 400 l/mm grating on the 
 GMOS instrument on Gemini. We have made an approximately relative flux 
 calibration of the spectrum based on the instrument throughput but have not
 attempted an absolute flux calibration. The exposure time was 4000s.  The 
 spectrum is shown in the rest frame of the host galaxy at a redshift
 $z=4.941$.  and the strong lines of L$\alpha$, SiII, OI, CII SiIV and CIV are
 marked. The shaded regions show the position of the atmospheric bands and
 strong night-sky lines. \label{fig1:spec}}
\end{center}
\end{figure}

We obtained a spectrum with the GMOS spectrograph on Gemini North,
commencing at 2006 May 10.547 UTC just over 2.5 hours after the
burst (Price et al. 2006b). Four 
1000s exposures were obtained with the 400 lines/mm grating
giving a resolution of 1900 and wavelength coverage from 5950~\AA\ to
10200~\AA.  Standard CCD reduction steps were performed with the
\texttt{gmos} package within IRAF, pairs of exposures were
sky-subtracted and combined, the spectra extracted using
\texttt{apextract}, and the two spectra summed to yield the final
product.
The spectrum was approximately relatively flux
calibrated using the instrument throughput. The final spectrum is
shown in Figure~\ref{fig1:spec}.

%
%

 The spectrum shows a damped L$\alpha$ line and an exensive set of metal lines
at a redshift of $z=4.941$ which we take to be the redshift of the host
galaxy. Some of the stronger features in the spectrum are marked on
Figure~\ref{fig1:spec}. The spectrum also shows a strong break across the
L$\alpha$ wavelength corresponding to the onset of the L$\alpha$ forest at
shorter wavelengths. The mean transmission in the L$\alpha$ forest is 18\%,
consistent with L$\alpha$ forest transmissions seen in quasar spectra at this
redshift (Songaila 2004).

\section{Host properties}
\label{sechost}

\begin{figure}
\begin{centering}
\includegraphics[viewport=20 0 446 626,width=2.4in,clip,angle=90,scale=0.85]{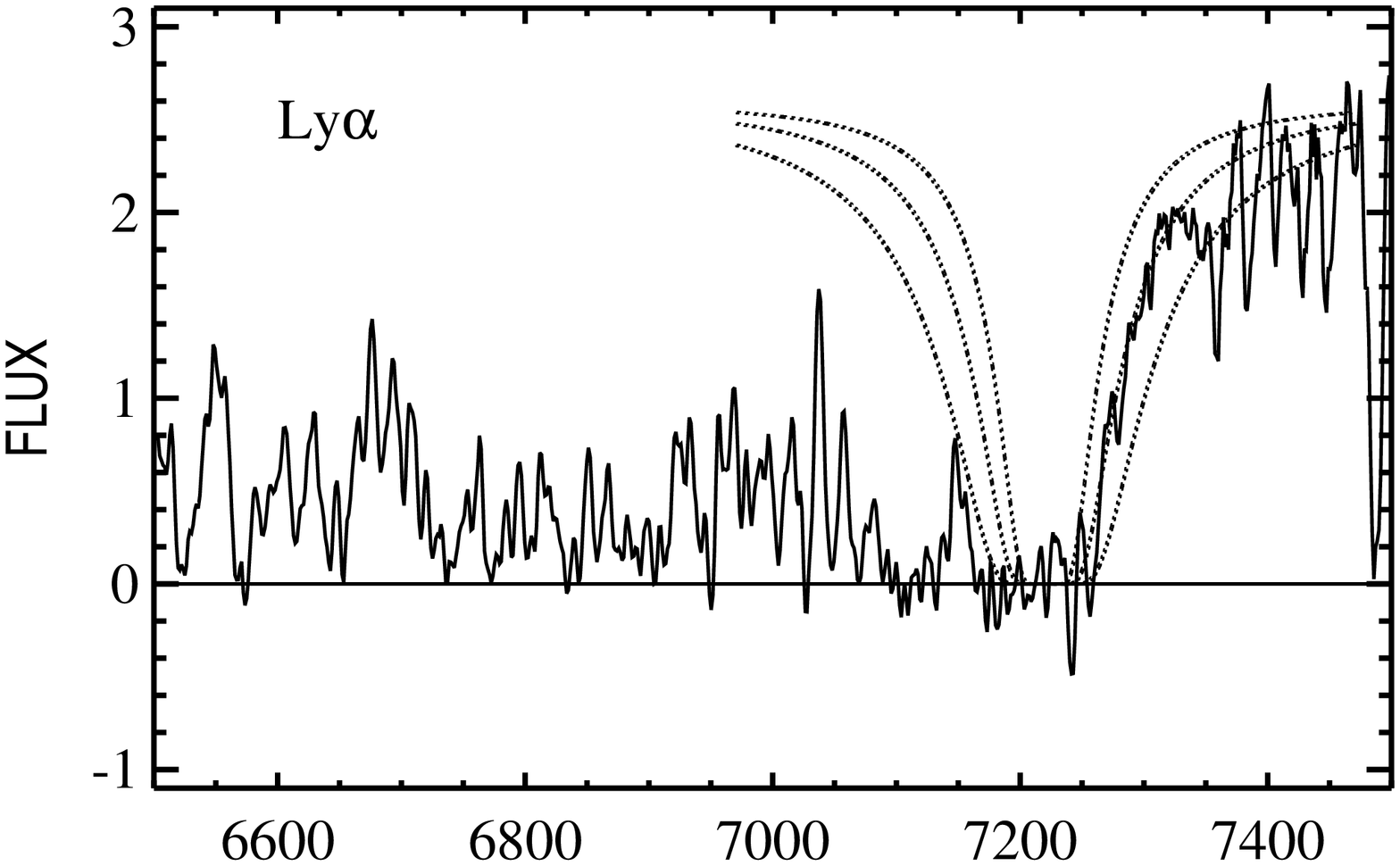}
\includegraphics[viewport=20 0 446 626,width=2.4in,clip,angle=90,scale=0.85]{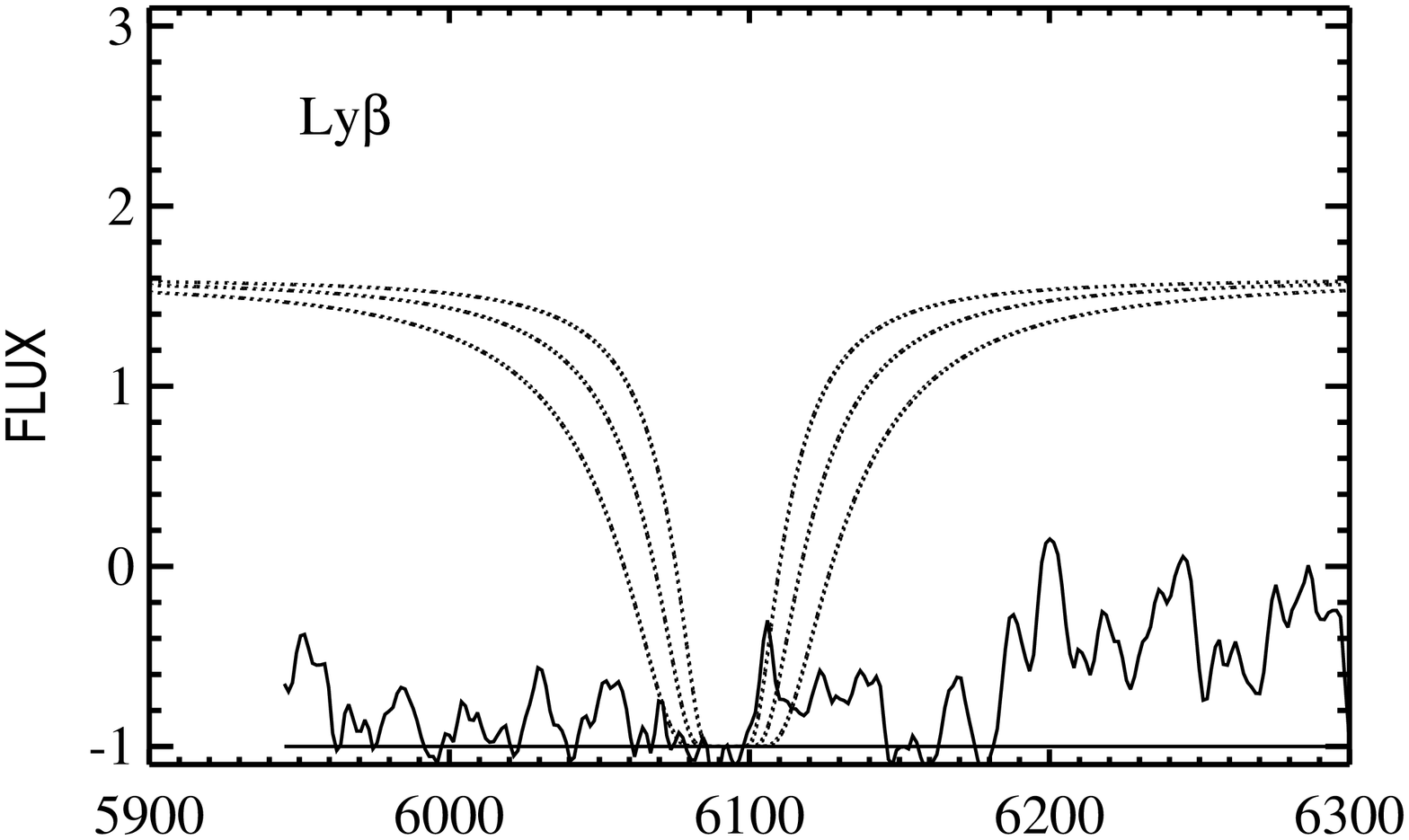}

  \caption{Damped L$\alpha$ and L$\beta$ profiles for N(H1) of
        $1\times10^{21}$ cm$^{-2}$, $2\times10^{21}$ cm$^{-2}$
        and $4\times10^{21}$ cm$^{-2}$.
  \label{fig2:damped}
  }
\end{centering}
\end{figure}

  We first measured a redshift of $z=4.941$ from the weaker singly ionized
metal lines in the spectrum.  Fits to the damped L$\alpha$ and L$\beta$ lines
centered at the wavelength corresponding to this redshift are shown in
Figure~\ref{fig2:damped}. The best fit to the red wing of the damped L$\alpha$
profile is given for a N(HI)=$1.7\times10^{21}$ cm$^{-2}$ and higher column
densities are prohibited. The L$\beta$ profile favors a slightly lower value
of N(HI)=$1.0\times10^{21}$ cm$^{-2}$ but this constraint is based on a single
pixel which could be contaminated.  We therefore adopt a range of
$\log N({\rm H1})=21.0-21.2$ for the logarithmic column density.

%
%

A second strong absorption feature seen in the L$\alpha$ forest at 7100~\AA\ is
not saturated at L$\beta$ and is not a DLA. This neutral hydrogen excess may
be caused by the higher density intergalactic medium in the vicinity of the
galaxy and it will be interesting to see if this is common in GRBs at these
redshifts since it could provide considerable infomation on the structure and
ionization of the overdense regions in which the galaxies are forming.

\begin{figure}
\begin{centering}
\includegraphics[viewport=20 0 486 376,width=3.0in,clip,angle=0,scale=0.85]{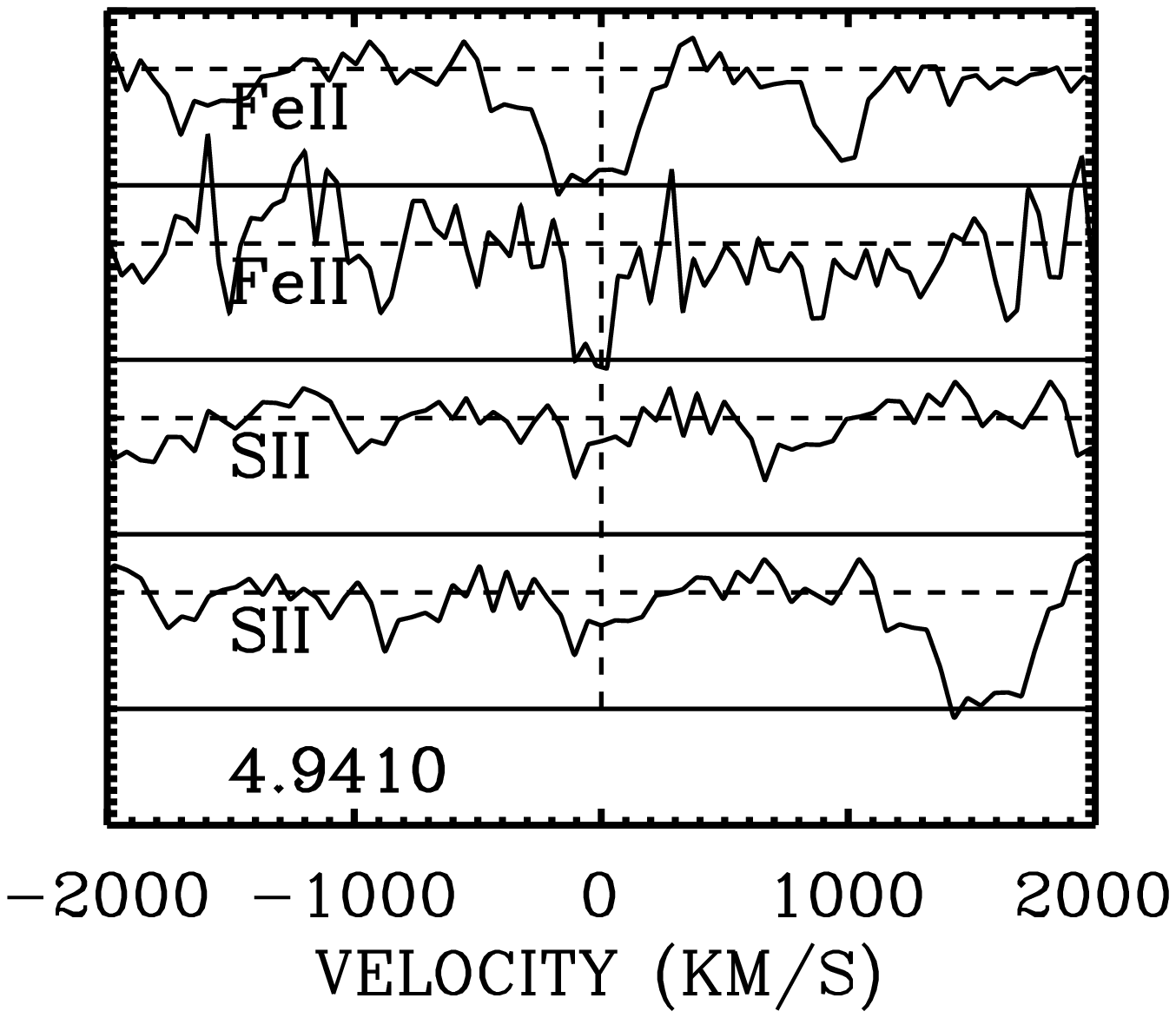}
\includegraphics[viewport=20 0 486 376,width=3.0in,clip,angle=0,scale=0.85]{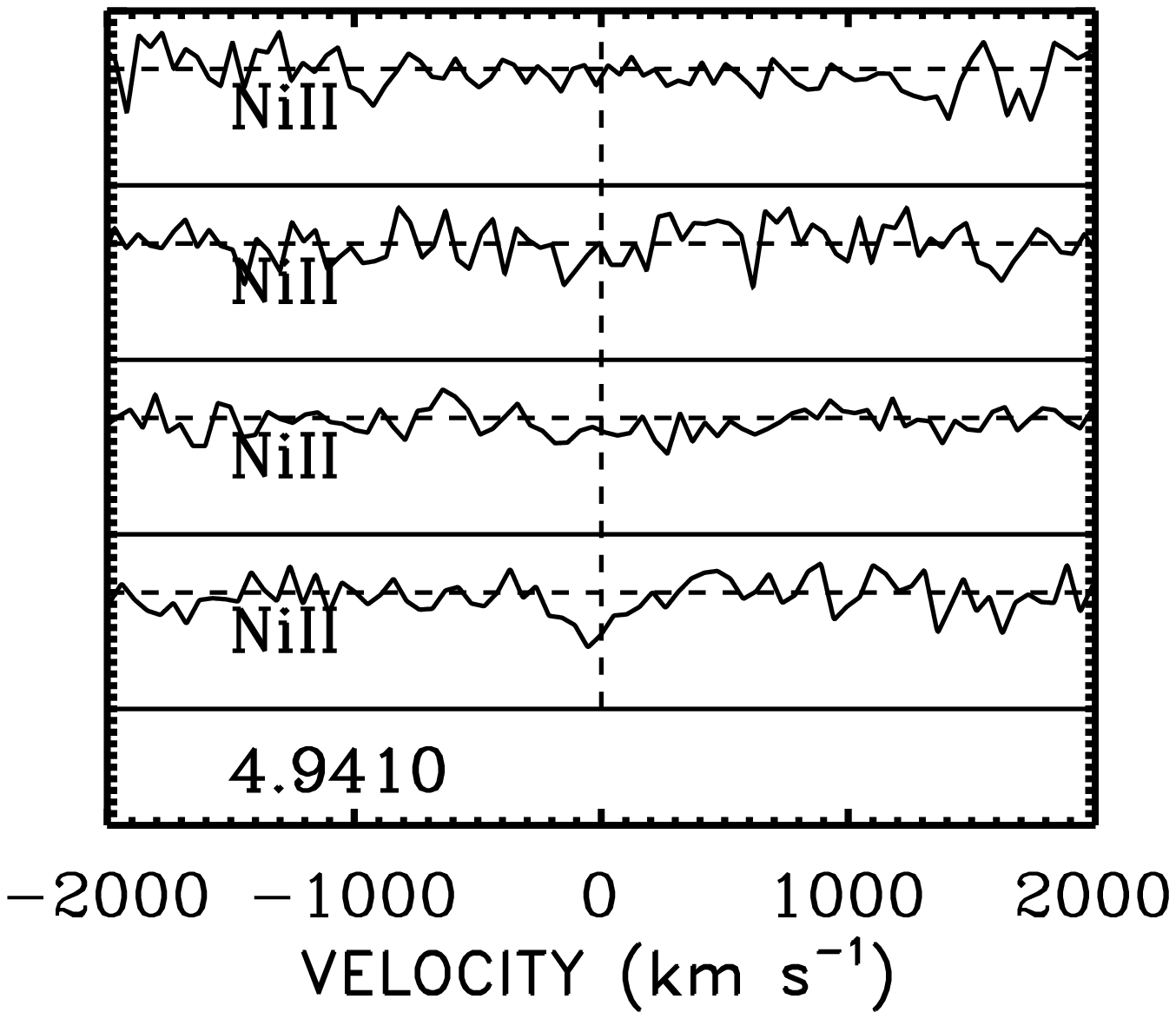}

  \caption{{\it Upper panel, bottom to top}:  lines of SII1251, SII1254,
          FeII1608, FeII1260.  {\it Lower panel, bottom to top}:  lines
          of NiII1317, NiII1370, NiII1455 and NiII1467.
          For each species the lines are shown in order of decreasing
          oscillator strength with the stronger lines at the bottom
          and the weaker at the top. The higher noise at the Fe1608
          line is a result of decreasing throughput at longer
          wavelengths.\label{fig3:weak}}
\end{centering}
\end{figure}

Although we detect many lines from the galaxy, most are strong
and, in low resolution spectra like the present one, the lower
limits on the column densities which can be obtained from 
such lines are not particularly useful. We therefore focus
on the weaker lines of Ni, S and Fe which are seen in the spectrum.
The absorption profiles of some of these lines are shown
in Figure~\ref{fig3:weak}.

Even in these moderately weak lines saturation can be a problem and saturated
components can be masked by unsaturated features in the lines. This problem,
long familiar in interstellar medium studies, has recently been re-emphasized
by Prochaska (2006) where an extensive discussion and critique of GRB host
metallicity measurements can be found, as well as the many historical
references on the topic. For the present work we measured the various Ni, S
and Fe lines under the assumption that they are unsaturated and lie in the
strictly linear portion of the curve of growth. The column densities measured
in this way and their one sigma uncertainties are given in Table~1.  We
emphasize that these are only lower limits to the column density. If there are
narrow saturated components concealed in the absorption lines the column
density could be higher.  However, even these lower limits already place a
lower limit on the metallicity of about $-1$ relative to solar
(Asplund et al.~\cite{asplund} ; Table~2)
which is already considerably higher than the values seen in DLAs at $z>4$
(Prochaska 2003).  We illustrate this in Figure~\ref{fig4:metalz}.


\begin{deluxetable}{llll}
\small
\tablewidth{230pt}
\tablecaption{Column Densities\label{tbl:1}}
\tablehead{
\colhead{Ion} & \colhead{$\lambda$\ (\rm\AA)} & \colhead{$f$}  &
\colhead{$\log N\ ({\rm cm}^{-2})$}
}
\startdata
 \ion{H}{1} & 1215.67 & 0.416 & $21.1 \pm 0.1$ \nl
 \ion{Fe}{2} & 1608.45 & 0.062 & $> 15.9$       \nl
 \ion{S}{2}  & 1250.58 & 0.005 & 15.6 (14.8 -- 15.9) \nl
             & 1253.81 & 0.011 & $15.5 \pm 0.2$      \nl
 \ion{Ni}{2} & 1317.22 & 0.145 & $14.4 \pm 0.2$      \nl
             & 1370.13 & 0.131 & $14.0 \pm 0.2$      \nl
             & 1454.84 & 0.0595 & 14.4 (14.0 -- 14.6) \nl

\enddata
\tablecomments{All errors are $1\sigma$.}

\end{deluxetable}


\begin{deluxetable}{lll}
\small
\tablewidth{230pt}
\tablecaption{Metallicities\label{tbl:2}}
\tablehead{
\colhead{Element} & \colhead{Solar} & \colhead{Relative}
}
\startdata
 S   &  $-4.86$    &    $-0.8 \pm 0.3$  \nl
 Fe  &  $-4.55$    &    $-0.3$ to $-0.8$ \nl
 Ni  &  $-5.77$    &    $-1.2 \pm 0.3$  \nl

\enddata

\end{deluxetable}

 If the signal to noise is high enough we can use very weak lines to
measure upper bounds on the metallicity even in low resolution spectra
(e.g. Savaglio 2006). However, the quality of the very high redshift GRB
spectra is not generally adequate to do this. We tested the limits we
could obtain in the present data using the very weak Fe1611 and Ni1467
lines. In each case we created a model with two components each with
$b=8\ {\rm km\ s}^{-1}$ and two broader ($b=50\ {\rm km\ s}^{-1}$)
components and adjusted the column densities to fit the stronger
lines. (Here $b$\ is 0.60 times the full width at half maximum of the
absorption line (Cowie \& Songaila~\cite{cs86}).  We varied the
column densities in the narrow components and compared the resulting
Voigt profiles with the weak lines. Within the noise and continuum
fitting uncertainties we cannot obtain a robust upper limit on the
column densities.

 The spectrum also shows strong high ionization lines with $\log
N({\rm NV})$ in excess of 14.5 and $\log N({\rm CIV})$ and $\log
N({\rm SiIV})$ in excess of 15. Strong fine-structure lines are seen for
\ion{Si}{2}, \ion{C}{2} and \ion{O}{1}, as is common to all the GRB absorbers
(Vreeswijk et al.~\cite{vreeswijk04}; Berger et al.~\cite{berger}; Chen et
al.~\cite{chen}). Given the uncertainty in estimating the ground state column
densities we do not attempt to infer gas densities but it is clear, as has
been noted by many authors, that GRB absorption is produced in much higher gas
density enviroments than DLAs in quasar sightlines, which only extremely
rarely show such lines (e.g. Chen et al.~\cite{chen}).
\section{Discussion}
\label{secdisc}
\begin{figure}
\begin{centering}
\includegraphics[viewport=20 0 486 646,width=2.8in,clip,angle=90,scale=0.85]{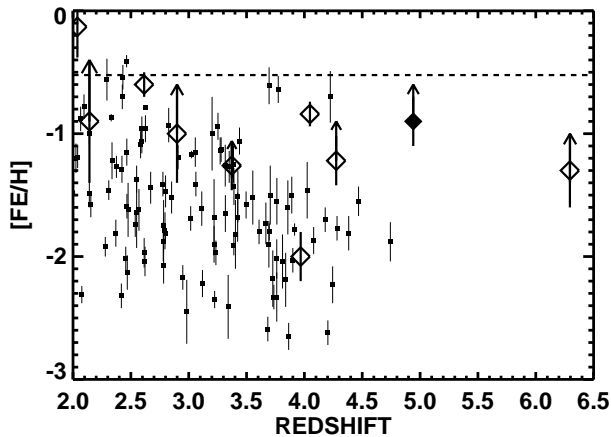}

  \caption{Metallicity relative to solar versus redshift for the GRB
host galaxies (diamonds) superimposed on the corresponding
measurements of metallicity in damped Lyman alpha systems along quasar
sightlines, taken from the compilation of Prochaska et al. (2003).
The value measured in this paper is shown as the filled diamond and is 
based on the SII, FeII, and NiII measurements of Table 2.  The open
diamonds show values from the literature (Savaglio et al. 2003,
Vreeswijk et al, 2004,2006 Chen et al. 2005, Starling et al. 2005,
Watson et al. 2005, Berger et al. 2006 and Fynbo et al. 2006). For 
lower resolution measurements (including the present measurement)
where saturation effects could significantly raise the measured
metallicity we plot the data with an upward pointing arrow.  The
dashed line shows the metallicity above which GRBS are not expected to
form in collapsar models.  \label{fig4:metalz} }
\end{centering}
\end{figure}

In Figure~\ref{fig4:metalz} we show all of the currently available
metallicity measurements for $z>2$ GRB host galaxies from the
references summarized in the figure caption. The present measurement
of GRB060510B is shown as the filled diamond with upward pointing
arrow to emphasize that it is strictly a lower limit. We have shown
other measurements of the GRB hosts as open diamonds. We show
measurements based on low resolution and moderate signal to noise
spectra with upward pointing arrows on $1~\sigma$ error bars while
values based on high resolution or high signal to noise observations
are shown with just the $1~\sigma$ errors.  Irrespective of redshift
the bulk of the GRB metallicities reported so far lie at or above
about a tenth solar. However, GRB hosts clearly have a wide range of
metallicities, for example, compare GRB050730 ($z=3.969$; Chen et
al. 2005) with a metallicity of $-2$ and GRB000926 ($z=2.038$;
Savaglio et al. 2003) with a value of $-0.13$.

The typical metallicity of GRB host galaxies is considerably higher than the
metallicities found in  DLAs, which are shown in Figure~\ref{fig4:metalz} as
the filled squares. However, there is only one $z>4.5$ DLA with a measured
metallicity (Songaila and Cowie 2002) so the comparison sample at the highest
redshifts is very limited.  This effect and the higher gas densities evidenced
by the fine structure lines are natural consequences of the selection biases.
GRB sightlines target star forming regions of galaxies where densities and
metallicities will be higher, whereas the cross section-weighted DLAs in
quasar sightlines probe more extended lower density and lower metallicity
regions in galaxies and may also be weighted to intrinsically lower luminosity
galaxies.

 Ultimately we would like to use GRBs to probe the star formation history of
the universe (Lamb and Reichert 2000, Price et al. 2006).  They have many
advantages for this purpose not the least of which is that they sample the
rate at which individual stars form irrespective of the mass or luminosity of
the host galaxy. We can therefore measure the total star formation over the
entire range of galaxies. However, before we can do this we must understand
the biases in the selection of the stars which become GRBs and determine
whether we can succesfully allow for such selection effects.

 The most probable bias is that GRBs occur only in low metallicity galaxies
and that we may therefore miss all of the star formation occuring in more
evolved systems.  In the collapsar models GRBs are formed in single massive
stars only if the metallicity is below $\sim$0.3 Z$_{\odot}$ (e.g.  Woosley
and Heger 2006). If such a bias is present it could introduce a strong
redshift dependence in the inferred star formation history if the average
metallicity of the host galaxies is lower at high redshift.

The evidence of Figure~\ref{fig4:metalz} is somewhat confusing as regards this
issue. In the first place it shows that while most metallicities in GRB hosts
could fall below 0.3 Z$_{\odot}$, at least one (GRB\,000926, $Z=0.7$
Z$_{\odot}$, Savaglio et al.~\cite{savaglio}) has a metallicity well above the
critical value. Fynbo et al. (2006) suggest that this indicates that
collapsars resulting from single massive stars are not the only progenitors of
long GRBs or that massive stars with Z $>$ 0.3 Z$_{\odot}$ can also produce
long GRBs. It is also possible that this particular GRB formed in a lower
metallicity region of the galaxy than that traversed by the
sight-line. However, if some of the measured metallicities are underestimates
this problem would become much more severe and we would have to conclude that
GRBs regularly formed in high metallicity environments. This issue clearly
requires a much larger sample of very high quality spectra to resolve.

 Furthermore if we do assume that GRBs are primarily formed in lower
metallicity stars then Figure~\ref{fig4:metalz} strongly suggests that
metallicity bias will be extremely important in determining star formation
rates.  Most of the measured values lie just below the critical value and
presumably as their metallicity continues to increase would drop from the
sample. Since a large fraction of the star formation would occur in the higher
metallicity objects we would miss much of the activity. With precise
metallicity measurements we could make a detailed comparison with models of
the star formation to see if the distribution was consistent and the
metallicity threshold was evident but such a comparison is premature
given the present quality of the data.

\acknowledgements
These data were obtained at the Gemini Observatory under Program ID
GN-2006A-Q-14.  We would like to thank Nestor Mirabal for helpful discussions
and the Gemini North observing team for carrying out the observations. This
work was supported by NSF grants AST04--07374 and AST06--07871, by NASA
grants HST--GO--10616.02--A and Swift NN G05 G40G, and by Spitzer contract
1282161.

\end{document}